\documentclass[letterpaper, 10 pt, conference]{ieeeconf}  

\IEEEoverridecommandlockouts                              

\usepackage{graphics} 
\usepackage{epsfig} 
\usepackage{mathptmx} 
\usepackage{times} 
\usepackage{amsmath} 
\usepackage{amssymb}  
\usepackage{bm} 
\usepackage{xcolor}
\usepackage{algorithm}
\usepackage{algpseudocode}
\usepackage{booktabs}
\usepackage{balance}

\algnewcommand\algorithmicinit{\textbf{Init:}}
\algnewcommand\Init{\item[\algorithmicinit]}

\title{\LARGE \bf
Energy system device sizing under closed-loop operation with imperfect forecasts
}

\author{Cara Koepele$^{1,2}$, Kevin Wallington$^{1}$, Georgios Mavromatidis$^{2}$, John Lygeros$^{1}$%
\thanks{$^{1}$Automatic Control Laboratory, ETH Z\"{u}rich, 
        8092 Z\"{u}rich, Switzerland,
        {\tt\small ckoepel@ethz.ch, kwallington@ethz.ch, jlygeros@ethz.ch}}%
\thanks{$^{2}$Urban Energy Systems Laboratory, EMPA,
        8600 D\"{u}bendorf, Switzerland,
        {\tt\small georgios.mavromatidis@empa.ch}}%
}

\begin{document}

\maketitle
\thispagestyle{empty}
\pagestyle{empty}

\begin{abstract}

The value of renewable energy devices depends on their operation, which must adapt to weather and demand using limited, imperfect forecasts that update over time. Conventional energy systems planning practices assume unlimited foresight in a single-level formulation, potentially leading to long-term investment in suboptimal device sizes. Accounting for realistic operation corresponds to a bi-level problem: the upper level selects device sizes, while the lower level contains closed-loop operation under evolving forecasts. We solve this problem using backpropagation MPC (BP-MPC), a gradient-based method, with a deterministic MPC as the lower-level controller. Using forecast errors generated by a parameterized martingale model of forecast evolution, we investigate how forecast accuracy affects the benefits of operation-aware energy systems planning and the resulting device sizes. In a case study of an energy system comprising photovoltaics, a battery, a heat pump, and thermal storage with uncertain weather and energy demands, accounting for closed-loop operation yields higher average cost savings as forecast error increases. The resulting designs achieve these savings by progressively shifting investment from battery storage to solar generation and thermal storage, which buffers forecast mismatches.

\end{abstract}

\section{Introduction}
The return on renewable energy investments depends on how the installed devices are operated under uncertain weather and energy demands. Standard energy systems planning methods typically co-optimize device sizes with an open-loop operating schedule for a lifetime of future disturbances \cite{ding_review_2022, gabrielli_optimal_2018, mavromatidis_design_2018}. This approach assumes that devices are dispatched with foresight that is unlimited in length and fixed in time. In practice, devices are dispatched by rule-based or receding horizon controllers \cite{smith_control_2023} that use forecasts with a limited horizon and errors. As a result, energy systems planning with unlimited foresight can yield suboptimal device sizes, increasingly so as poorer forecasts magnify differences between assumed and actual operation \cite{bylling_impact_2020, peterssen_impact_2024}.

Accounting for realistic operations during planning can be formulated as a bi-level problem: the upper level chooses device sizes to minimize lifetime cost, and the lower level contains a receding-horizon controller that determines power setpoints from the available forecasts at every timestep. This is an instance of control co-design \cite{garciasanz_control_2019, allison_multidisciplinary_2014}, in which the controller is fixed and the plant is sized for its closed-loop performance. Solving this bi-level problem is challenging because the lower level comprises many linked optimization problems that give rise to the system's closed-loop trajectory.
Baker et al. \cite{baker_energy_2017} approximate this structure for a wind energy system by summing the objectives of successive MPC horizons in a single storage-sizing problem, which can allow earlier decisions to depend on later information. Fortenbacher et al. \cite{fortenbacher_optimal_2018} instead adapt Benders decomposition for battery sizing and placement in low-voltage grids, using MPC simulations to update Benders cuts. Notably, neither approach establishes optimality for the closed-loop planning objective.

Other studies use zeroth-order optimization methods to solve the bi-level device-sizing problem \cite{ding_review_2022}. For example, Rullo et al. \cite{rullo_integration_2019} use a genetic algorithm to size devices in a renewable energy system with evaluations of MPC management under perfect forecasts. Sorourifar et al. \cite{sorourifar_computationally_2023} employ Bayesian optimization to size a PV--battery system under only two levels of forecast error. 
Nonetheless, derivative-free search becomes expensive as the number of design variables grows. Moreover, prior works have not characterized how the level of forecast error impacts the added value of operation-aware design.

First-order methods offer an alternative by using gradients of closed-loop performance with respect to the optimized parameters. Although constrained MPC policies can be nonsmooth, their solution maps can be differentiated through the optimality conditions almost everywhere under suitable regularity assumptions \cite{amos_differentiable_2018}, and the resulting sensitivities can be propagated along a closed-loop trajectory \cite{agrawal_learning_2020}. This concept has been used to select MPC parameters by taking gradient steps based on the closed-loop cost, as in reinforcement learning-based MPC tuning \cite{gros_data-driven_2020} or backpropagation MPC (BP-MPC) \cite{zuliani_bp-mpc_2025}. By treating device capacities as parameters of the MPC optimization problem, these approaches can be extended to energy systems design.

This paper makes two contributions. First, we extend BP-MPC to plant design by differentiating the closed-loop cost with respect to device capacities through both the MPC and a low-level controller that adjusts its inputs to the realized disturbances. Second, we use this method to quantify how forecast error affects optimal designs and annualized costs, comparing operation-aware and single-level sizing under the same MPC across forecast error levels. The system in this study comprises photovoltaics (PV), a battery (BESS), a heat pump (HP), and short-term thermal energy storage (TES). Forecast errors are varied using a parameterized martingale model of forecast evolution (MMFE) \cite{heath_modeling_1994}. Our results show that the average cost advantage of BP-MPC sizing grows with the standard deviation of the one-step forecast error, while increasing forecast error shifts investment from the BESS toward PV and TES.

\section{Energy System Model}\label{sec:model}

We consider a multi-energy system comprising PV, BESS, HP, and TES, employing similar models as in \cite{behrunani_distributed_2024, micheli_stochastic_2023}. The PV generates electrical power $P^\mathrm{pv}_t$ proportional to the installed capacity $p^{\mathrm{pv}}$ and a capacity factor $\eta^\mathrm{pv}_t$:
\begin{equation}\label{cst:pv}
    P^\mathrm{pv}_t = \eta^\mathrm{pv}_t \; p^\mathrm{pv}.
\end{equation}
Here, the capacity factor is determined from ambient temperature and global irradiance \cite{skoplaki_temperature_2009}.

The HP has a capacity of $p^\mathrm{hp}$ and converts electrical power $P^{\mathrm{hp}}_t$ to thermal power $Q^\mathrm{hp}_t$ with a constant COP, $\eta^{\mathrm{COP}}$:
\begin{subequations}\label{cst:hp}
\begin{align}
    &Q^{\mathrm{hp}}_t = \eta^{\mathrm{COP}} P^{\mathrm{hp}}_t,\\
    &Q^\mathrm{hp}_t \in [0,\; p^\mathrm{hp}].
\end{align}
\end{subequations}

The BESS storage dynamics and constraints are:
\begin{subequations}\label{cst:bess}
\begin{align}
    &E^{\mathrm{bess}}_{t+1} = \eta^{\mathrm{bess}}E^{\mathrm{bess}}_{t} + \Delta t(\eta^{\mathrm{bess}}_{\mathrm{ch}} P^{\mathrm{ch}}_t - \frac{1}{\eta^{\mathrm{bess}}_{\mathrm{dc}}} P^{\mathrm{dc}}_t), \label{eq:bess_dyn_cst}\\
    &E^{\mathrm{bess}}_{t} \in [ \underline{a}^\mathrm{bess},\; \overline{a}^\mathrm{bess}]p^{\mathrm{bess}},\label{eq:bess_nrg_cst}\\
    &P^{\mathrm{ch}}_t, P^{\mathrm{dc}}_t \in [ 0,\; \overline{b}^{\mathrm{bess}} ] p^{\mathrm{bess}}, \label{eq:bess_pwr_cst}
\end{align}   
\end{subequations}
where $E^{\mathrm{bess}}_{t}$ is the state-of-charge, with $P^{\mathrm{ch}}_t$ and $P^{\mathrm{dc}}_t$ as the charge and discharge power, respectively, and $\Delta t$ is the timestep. The BESS has a capacity of $p^{\mathrm{bess}}$, storage efficiency of $\eta^{\mathrm{bess}}\in(0,1)$, and charge and discharge efficiencies of $\eta^{\mathrm{bess}}_{\mathrm{ch}}\in(0,1)$ and $\eta^{\mathrm{bess}}_{\mathrm{dc}}\in(0,1)$, respectively. Additionally, the BESS state of charge is limited to be between $\underline{a}^\mathrm{bess}$ and $\overline{a}^\mathrm{bess}$, and the charge/discharge powers to a fraction $\overline{b}^{\mathrm{bess}}$ of the BESS capacity.

The TES storage dynamics and constraints are:
\begin{subequations}\label{cst:tes}
\begin{align}
    &E^{\mathrm{tes}}_{t+1} = \eta^{\mathrm{tes}}E^{\mathrm{tes}}_{t} + \Delta t(\eta^{\mathrm{tes}}_{\mathrm{ch}} Q^{\mathrm{ch}}_t - \frac{1}{\eta^{\mathrm{tes}}_{\mathrm{dc}}} Q^{\mathrm{dc}}_t) \label{eq:tes_dyn_cst},\\
    &E^{\mathrm{tes}}_{t} \in [ \underline{a}^\mathrm{tes}, \; \overline{a}^\mathrm{tes}]p^{\mathrm{tes}}\label{eq:tes_nrg_cst},\\
    &Q^{\mathrm{ch}}_t, Q^{\mathrm{dc}}_t \in [ 0, \; \overline{b}^{\mathrm{tes}} ] p^{\mathrm{tes}},\label{eq:tes_pwr_cst}
\end{align}   
\end{subequations}
where the nomenclature follows that of the BESS equations, and $Q$ is used for thermal charge and discharge. 

The energy system is connected to the electric grid and can purchase electricity $P^{\mathrm{buy}}_t$ at time-of-use tariffs $c^{\mathrm{buy}}_t \in\{\underline{c}^{\mathrm{buy}},\overline{c}^{\mathrm{buy}}\}$ and sells excess electricity $P^{\mathrm{sell}}_t$ at a predetermined price $c^{\mathrm{sell}}_t$. The grid exchange and net cost satisfy
\begin{subequations}\label{cst:grid}
\begin{align}
    &P^{\mathrm{buy}}_t, P^{\mathrm{sell}}_t \geq 0,\\
    &C^{\mathrm{grid}}_t = c^{\mathrm{buy}}_t P^{\mathrm{buy}}_t - c^{\mathrm{sell}}_t P^{\mathrm{sell}}_t.
\end{align}   
\end{subequations}
The purchase price of electricity is always assumed to be higher than the sale price, i.e., $c^{\mathrm{buy}}_t > c^{\mathrm{sell}}_t,\;\forall t$.
There is no heat grid considered in this study, so a mismatch in heat supply and demand results in heat violations, denoted as $Q^{-}_t\geq 0$ and $Q^{+}_t\geq0$ for unmet and excess heat, respectively. 

Finally, the electric demand $P^\mathrm{L}_t$ and heat demand $Q^\mathrm{L}_t$ must be met through the nodal energy balances:
\begin{subequations}\label{cst:nrg_bals}
\begin{align}
    &P^{\mathrm{L}}_t -P^\mathrm{pv}_t = P^{\mathrm{hp}}_t - P^{\mathrm{ch}}_t + P^{\mathrm{dc}}_t + P^{\mathrm{buy}}_t - P^{\mathrm{sell}}_t,\\
    &Q^{\mathrm{L}}_t = Q^{\mathrm{hp}}_t - Q^{\mathrm{ch}}_t + Q^{\mathrm{dc}}_t + Q^{-}_t - Q^{+}_t.
\end{align}
\end{subequations}

For the entire energy system, the states are the stored energies $x$, the decision variables are the component power setpoints $u$, and the disturbances $\omega$ are the PV generation and energy demands. The design decision $p$ comprises all device capacities in the system. In summary,
\begin{align*}
    x_t&=[E^{\mathrm{bess}}_t,\; E^{\mathrm{tes}}_t]^\top,\\
    u_t&=[P^{\mathrm{hp}}_t,\; P^{\mathrm{ch}}_t,\; P^{\mathrm{dc}}_t,\; P^{\mathrm{buy}}_t,\; P^{\mathrm{sell}}_t,\; Q^{\mathrm{ch}}_t,\; Q^{\mathrm{dc}}_t,\; Q^{\mathrm{-}}_t,\; Q^{\mathrm{+}}_t]^\top,\\
    \omega_t&=[P^{\mathrm{L}}_t,\; Q^{\mathrm{L}}_t,\; \eta^\mathrm{pv}_t]^\top,\\ 
    p &= [p^{\mathrm{pv}},\;p^{\mathrm{bess}},\;p^{\mathrm{hp}},\;p^{\mathrm{tes}}]^\top.
\end{align*}

\section{Energy system operation with deterministic MPC}
The energy system model described in Section \ref{sec:model} is embedded in a deterministic MPC controller that dispatches the devices to meet electrical and thermal demands. Given installed device capacities $p$, the controller minimizes operating costs and penalties for heat violations. At each timestep $t$, we assume that the state $x_t$ is measured and forecasts $\hat{\omega}_{t|t+k}$ of disturbance $\omega_{t+k}$ are available over a prediction horizon of $N$ timesteps, with $k=0,\ldots,N-1$. To determine the next control action, the MPC solves the following optimization problem:
\begin{subequations}\label{opt:mpc}
\begin{align}
    \underset{\hat{x},\hat{u}}{\text{argmin}}\;&\sum_{k=0}^{N-1} \hat{C}^{\mathrm{grid}}_k (\hat{u}_k) + c^{\mathrm{viol}}(\hat{Q}^{-}_k + \hat{Q}^{+}_k) + C^{\mathrm{reg}}(\hat{x},\hat{u}),\\
    \text{s.t. }& \hat{x}_{k+1} = A\hat{x}_k + B\hat{u}_k,\;\label{opt:mpc_dyn_cst}\\ 
    &H_{\mathrm{eq}} \hat{u}_k = h_{\mathrm{eq}}(\hat{\omega}_{t|t+k}, p),\label{opt:mpc_eq_cst}\\ 
    &H_x \hat{x}_k \leq h_x(p),\;H_u \hat{u}_k \leq h_u(p),\label{opt:mpc_dev_cst}\\
    &\forall k=0,...,N-1,\;\hat{x}_0 = x_t. \label{opt:mpc_init_cst}
\end{align}   
\end{subequations}
The decision variables of \eqref{opt:mpc} are the predicted state trajectory $\hat{x}=(\hat{x}_0,\ldots,\hat{x}_N)$ and control sequence $\hat{u}=(\hat{u}_0,\ldots,\hat{u}_{N-1})$. 
The constraints follow from the energy system model in \eqref{cst:pv}--\eqref{cst:nrg_bals} and encode the storage dynamics \eqref{opt:mpc_dyn_cst}, energy balances \eqref{opt:mpc_eq_cst}, and grid and device constraints \eqref{opt:mpc_dev_cst}. The linear penalty coefficient $c^{\mathrm{viol}}$ represents the cost per unit of heat violation, which can be interpreted as the cost of covering mismatches with an auxiliary heater plus a discomfort premium. The cost function also includes the regularization term
\begin{equation}\label{eqn:reg}
   C^{\mathrm{reg}}(x,u) = \gamma\left(\lVert S^{-1}_x x\rVert^2_2 + \lVert S^{-1}_u u\rVert^2_2+d^\top S^{-1}_u u\right), 
\end{equation}
where $\gamma>0$ controls the regularization strength and $d$ assigns additional costs to discourage unnecessary energy throughput from simultaneous charging and discharging that could otherwise reduce the quadratic state penalty. Diagonal matrices $S_x$ and $S_u$ scale $x$ and $u$, respectively, so that the penalties are comparable across components. Together with the other linear costs, the assumed grid tariffs, and conversion losses, these terms discourage simultaneous opposing flows in storage and grid exchange, as well as unnecessary heat violations.

At each timestep $t$, \eqref{opt:mpc} is solved, and the first optimal control action defines the MPC proposal $\hat{u}_t := \hat{u}^{\ast}_0$. Forecast errors may cause this proposal to violate the actual energy balances. We therefore assume that a low-level controller adjusts it to a feasible input through the mapping $u_t=\Lambda(x_t,\hat{u}_t,\omega_t,p)$, which accounts for the actual disturbance through real-time balancing adjustments.
Mismatches at the electricity node are compensated through the grid, while those at the heat node are first compensated through the TES, with any remaining mismatch recorded as a heat violation once TES constraints are reached. Thus, MPC directly sets the HP and BESS setpoints, while grid purchases, TES setpoints, and heat violations follow from the actual energy balances. These rules make $u_t$ piecewise affine in $x_t$ and $\hat{u}_t$.

\section{Single-level device sizing}
We introduce a conventional single-level sizing formulation as a benchmark for operation-aware design. It jointly optimizes device capacities and an open-loop operating schedule under perfect foresight over a representative year. The resulting capacities are subsequently evaluated under the same MPC used to evaluate the bi-level design.

The formulation minimizes annualized investment and operating costs over a representative year of $T$ timesteps, which is much longer than the MPC horizon, $N$. We assume investment costs are linear in the capacities, $p$, with annualized per-unit costs $c^{\mathrm{inv}}$. Capacities satisfy the box constraints $p\in[\underline{p},\overline{p}]$, with a small $\underline{p}>0$ to avoid degenerate storage constraints. For simplicity, we neglect space considerations and limitations to discrete sizes available by manufacturers. Storage is installed at its minimum state of charge, so the initial state is $\underline{x}(p) := [\underline{a}^{\mathrm{bess}}p^{\mathrm{bess}}, \underline{a}^{\mathrm{tes}}p^{\mathrm{tes}}]^\top$. The capacities, state trajectory $x=(x_0,\ldots,x_T)$, and power setpoints $u=(u_0,\ldots,u_{T-1})$ are jointly determined by the following QP:
\begin{subequations}\label{opt:open_loop_design}
\begin{align}
    \underset{p,x,u}{\text{min}}\;&c^{\mathrm{inv}^\top}p + \sum_{t=0}^{T-1} C^{\mathrm{grid}}_t(u_t) + c^{\mathrm{viol}}(Q^{-}_t + Q^{+}_t)\\ &+ C^{\mathrm{reg}}(x,u),\nonumber\\
    \text{s.t. }& x_{t+1} = Ax_t + Bu_t,\label{opt:olc_dyn_cst}\\ 
    &H_{\mathrm{eq}} u_t = h_{\mathrm{eq}}(\omega_{t}, p),\label{opt:olc_eq_cst}\\ 
    &H_x x_t \leq h_x(p),\;H_u u_t \leq h_u(p),\label{opt:olc_dev_cst}\\
    &\forall t=0,...,T-1,\;x_0 = \underline{x}(p), p \in [\underline{p}, \overline{p}]. \label{opt:olc_init_cst}
\end{align}   
\end{subequations}
Problem \eqref{opt:open_loop_design} uses the realized disturbances, $\omega$, for the entire representative year over which perfect, unlimited foresight is assumed. Its optimal objective provides a lower bound on the objective achievable under MPC. To assess the resulting design under realistic operation, we fix its capacities, simulate the MPC in \eqref{opt:mpc}, and compute the annualized investment and realized operating costs. This closed-loop cost is the benchmark that bi-level sizing seeks to improve.

\section{Bi-level device sizing}

\subsection{The bi-level problem formulation}
We assume that the system is operated with a deterministic MPC under imperfect forecasts, giving rise to the bi-level sizing problem. 
\begin{subequations} \label{opt:bi-level}
\begin{align}
    \underset{p,x,u}{\text{min}}\;&c^{\mathrm{inv}^\top}p +\sum_{t=0}^{T-1} C^{\mathrm{grid}}_t(u_t) + c^{\mathrm{viol}}(Q^{-}_t + Q^{+}_t) \label{opt:bi-level-obj}\\ &+ C^{\mathrm{reg}}(x,u),\nonumber\\
    \text{s.t. }& x_{t+1} = Ax_t + Bu_t,\label{opt:bi-level-dyn}\\
    & u_t = \Lambda \left(x_t, \pi^{\mathrm{MPC}}(x_t,\hat{\omega}_t,p), \omega_t, p \right),\\
    &\forall t=0,...,T-1,\;x_0 = \underline{x}(p),\;p \in [\underline{p}, \overline{p}].
\end{align}   
\end{subequations}
Here, the control action $\hat{u}_t = \pi^{\mathrm{MPC}}(x_t, \hat{\omega}_t,p)$ is the first element of the decision trajectory, $\hat{u}$, computed from \eqref{opt:mpc}, and $u_t = \Lambda(x_t,\hat{u}_t,\omega_t,p)$ is the projection operator which ensures that input constraints are respected despite errors between the forecast, $\hat{\omega}_t$, and the realization, $\omega_t$. Because the input and state constraints are enforced by $\Lambda$ and $\pi^{\mathrm{MPC}}$, only the system dynamics, \eqref{opt:bi-level-dyn}, have to be imposed as constraints in the upper level of the problem. Note that problem \eqref{opt:bi-level} is posed for a single forecast error realization and thus approximates the expected closed-loop cost with one sample. Minimizing the expectation directly is left to future work.

\subsection{Solving the bi-level problem with Backpropagation MPC}
Solving \eqref{opt:bi-level} with gradient-based methods is challenging because the MPC policy $\pi^{\mathrm{MPC}}$ and the map $\Lambda$ are not differentiable everywhere. BP-MPC \cite{zuliani_bp-mpc_2025} is a projected gradient method developed to address this challenge, which can be shown to converge to a critical point of \eqref{opt:bi-level}. Since \eqref{opt:bi-level} is not convex, the limit point may not be the global optimum in general. BP-MPC replaces smoothness requirements with path-differentiability. Locally Lipschitz, definable functions, such as piecewise-affine ones, are path-differentiable \cite[Lemma~2]{zuliani_bp-mpc_2025} and admit a conservative Jacobian that obeys the chain rule \cite[Lemma~1]{zuliani_bp-mpc_2025}. We write $\frac{d(\cdot)}{dp}$ for elements of these conservative Jacobians and refer to them as sensitivities.

We define $\phi$ to be the upper-level objective of \eqref{opt:bi-level}. By the chain rule, the gradient of $\phi$ is
\begin{equation}\label{eq:grad}
  \frac{d\phi}{dp} = \frac{\partial\phi}{\partial p}
  + \sum_{t=0}^{T-1}\left( \frac{\partial\phi}{\partial x_{t+1}}\frac{dx_{t+1}}{dp}
  + \frac{\partial\phi}{\partial u_t}\frac{du_t}{dp} \right),
\end{equation}
where the state sensitivities follow the dynamics,
\begin{equation}\label{eq:dx_dp}
  \frac{dx_{t+1}}{dp} = A\frac{dx_t}{dp} + B\frac{du_t}{dp},
\end{equation}
initialized with $\frac{dx_0}{dp} = \frac{\partial \underline{x}(p)}{\partial p}$. The action sensitivities are
\begin{equation}\label{eq:du_dp}
  \frac{du_t}{dp} = \frac{\partial\Lambda}{\partial x_t}\frac{dx_t}{dp}
  + \frac{\partial\Lambda}{\partial\pi^{\mathrm{MPC}}}\left[
  \frac{\partial\pi^{\mathrm{MPC}}}{\partial x_t}\frac{dx_t}{dp}
  + \frac{\partial\pi^{\mathrm{MPC}}}{\partial p}\right]
  + \frac{\partial\Lambda}{\partial p}.
\end{equation}
Recall that $\Lambda$ is piecewise affine by design. The MPC policy is path-differentiable in $(x_t, p)$ if the MPC problem is feasible, strongly convex, and satisfies the linear independence constraint qualification (LICQ) \cite[Theorem~1]{zuliani_bp-mpc_2025}. Its conservative
Jacobian is then obtained by implicit differentiation of the dual problem. In our formulation, feasibility holds because grid exchange and heat violations are unconstrained, while strong convexity can be ensured by selecting $\gamma > 0$. We assume LICQ holds at the optimal solutions considered. Positive lower storage bounds $\underline{a}^{\mathrm{bess}}>0$ and $\underline{a}^{\mathrm{tes}}>0$ exclude empty-storage degeneracies, while strictly positive demands exclude zero-demand degeneracies. Beyond this, LICQ failures may arise from coincidental disturbance values but are considered unlikely.

Algorithm \ref{alg:BP-MPC} summarizes the energy systems planning variant of BP-MPC. At each iterate $p_i$, we simulate closed-loop operation over the planning horizon using forecasts and true disturbances, then compute the design gradient using \eqref{eq:grad}, \eqref{eq:dx_dp}, and \eqref{eq:du_dp}. A projected step updates the design, with step sizes satisfying $\sum_i\alpha_i=\infty$ and $\sum_i\alpha_i^2<\infty$. To damp oscillations, we smooth the gradients with a bias-corrected exponential moving average \cite{kingma_adam_2015}. For numerical conditioning, we also scale gradients using a diagonal matrix $D$ and clip their norm to $r$ via $\mathrm{clip}_r(g)=g\cdot\min\{1,r/\lVert g\rVert_2\}$. Note that these practical modifications are not covered by the convergence result of \cite{zuliani_bp-mpc_2025}.

\begin{algorithm}[!t]
\caption{BP-MPC for energy system device sizing}
\label{alg:BP-MPC}
\begin{algorithmic}[1]
\Require $p_0$, $x_0$, $\frac{dx_0}{dp}$, $\{\alpha_i\}_{i\in\mathbb{N}}$, $\beta$, $D$, $r$, $\omega$, $\hat{\omega}$
\Init $i \gets 0,\; m_0 \gets 0$
\While{not converged}
  \For{$t = 0,\dots,T-1$}
    \State Solve \eqref{opt:mpc} to get $\hat{u}_t = \pi^\mathrm{MPC}(x_t, \hat{\omega}_t, p_i)$
    \State $u_t \gets \Lambda(x_t, \hat{u}_t, \omega_t, p_i)$
    \State $x_{t+1} \gets A x_t + B u_t$
    \State Compute $\frac{du_t}{dp}$ and $\frac{dx_{t+1}}{dp}$ with \eqref{eq:du_dp} and \eqref{eq:dx_dp}
  \EndFor
  \State Compute $\frac{d\phi}{dp}$ with \eqref{eq:grad}
  \State $g_i \gets \mathrm{clip}_r\!\left(D \frac{d\phi}{dp}\right)$
  \State $m_{i+1} \gets \beta m_i + (1-\beta)\, g_i$
  \State $p_{i+1} \gets \mathbb{P}_{[\underline{p},\overline{p}]}\!\left[p_i - \alpha_i \frac{m_{i+1}}{1-\beta^{i+1}}\right]$
  \State $i \gets i + 1$
\EndWhile
\State \Return $p^\ast = p_i$
\end{algorithmic}
\end{algorithm}

\subsection{Forecast error modeling with MMFE}
We adopt the MMFE to generate evolving synthetic forecasts \cite{heath_modeling_1994} without tying their error behavior to a specific regression model. This framework has been widely used in supply chain management and applied in several energy management studies \cite{liu_optimal_2020, zhu_short-term_2020}. The MMFE describes forecast evolution through successive updates:
\begin{equation}
    \Delta \hat{\omega}_{t|t+i} = \hat{\omega}_{t+1|t+i} - \hat{\omega}_{t|t+i},
\end{equation}
where $\hat{\omega}_{t|t+i}$ is the forecast issued at time $t$ for the disturbance at timestep $t+i$.  

Forecasts of the disturbances at $t+i$ are updated at each timestep from $t+i-N+1$ until the timestep $t+i+1$, where N is the forecasting horizon, which is the same as the control horizon in \eqref{opt:mpc}. At timestep $t+i+1$, the true value $\omega_{t+i}$ is uncovered. The error in the forecast $e_{t|t+i}$ is the sum of all forecast updates from time $t$ until time $t+i$:
\begin{equation} \label{eq:mmfe_error}
    e_{t|t+i} = \sum_{j=0}^{i} \Delta \hat{\omega}_{t+j|t+i}.
\end{equation}

The vector of forecast updates $\Delta \hat{\Omega}_t$ over the prediction horizon $N$ is assumed to be drawn from an unbiased, stationary multivariate Gaussian distribution:
\begin{equation} \label{eq:mmfe_update_def}
    \Delta \hat{\Omega}_t = \begin{bmatrix}\Delta \hat{\omega}_{t|t} & \Delta \hat{\omega}_{t|t+1} & \cdots & \Delta \hat{\omega}_{t|t+N-1}\end{bmatrix} \sim \mathcal{N}(0, \Sigma),
\end{equation}
where $\Sigma$ is the covariance matrix of the forecast update vector. Following \cite{ma_scenario_2013}, we parameterize the covariance matrix with the variance of each forecast update $\sigma^2$ and error correlation $\rho\in(-1,1)$ as
\begin{equation} \label{eq:mmfe_param}
    \Sigma_{i,j}(\sigma, \rho) = \sigma^2 \rho^{\lvert i-j \rvert}.
\end{equation}
Higher $\sigma$ indicates higher forecast error overall. Positive $\rho$ increases the likelihood that updates have the same sign, whereas negative $\rho$ supports the converse. The forecasts of the disturbances in this study are assumed to be correlated over several hours, indicating a positive $\rho$. Notably, since we are sweeping over the aforementioned parameters, we do not fit this model to real forecasting data, which is left to future work.

Since $\Sigma$ is positive definite for $\lvert \rho \rvert < 1$, it can be decomposed as $\Sigma = V\cdot V^\top$ through the Cholesky decomposition. A sample of the forecast updates can be obtained as:
\begin{equation} \label{eq:mmfe_sample}
    \Delta \hat{\Omega}_t = [z_0, z_1,...,z_{N-1}] \cdot V^\top,
\end{equation}
where $z_i$ are independent standard normal variables. 

We independently generate normalized forecast errors for each disturbance from the multivariate Gaussian distribution defined in \eqref{eq:mmfe_update_def} and \eqref{eq:mmfe_param}, with identical $\sigma$ and $\rho$ and no cross-correlation between disturbances. Though forecasters for each disturbance may have different structures in practice, we choose to use common parameters to facilitate systematic comparisons. Errors are then scaled by a disturbance-specific, hourly reference value to promote realistic forecast values, such as lower irradiance at sunrise and sunset. The forecast is therefore:
\begin{equation}\label{eq:mmfe_error_scaled}
    \hat{\omega}_{t|t+i} = \max\{0, \omega_{t+i} - s^\top(h_{t+i})e_{t|t+i}\},
\end{equation}
where $s$ is the scaling vector based on the hour of the day the forecast is for, $h_{t+i}$. We scale PV capacity factor errors by a seasonal ``clear sky" trajectory while heat and electricity demand errors are scaled by a seasonal mean trajectory and a weekday/weekend trajectory, respectively. The forecast errors are clipped to prevent unrealistic negative forecasts. To generate all forecast errors, we sample the forecast update vector at every timestep $t=0,...,T+N$ and apply equation \eqref{eq:mmfe_error_scaled}.

Note that the MMFE is a generative error model that constructs errors from future updates, rather than a reproduction of real forecasting procedures. By sampling updates instead of errors, it yields forecasts that are coherent across issuances and unbiased before clipping, with an error spread that grows with lead time as updates accumulate.

\section{Results}
\begin{figure}[b]
    \centering
    \includegraphics[width=1.0\linewidth]{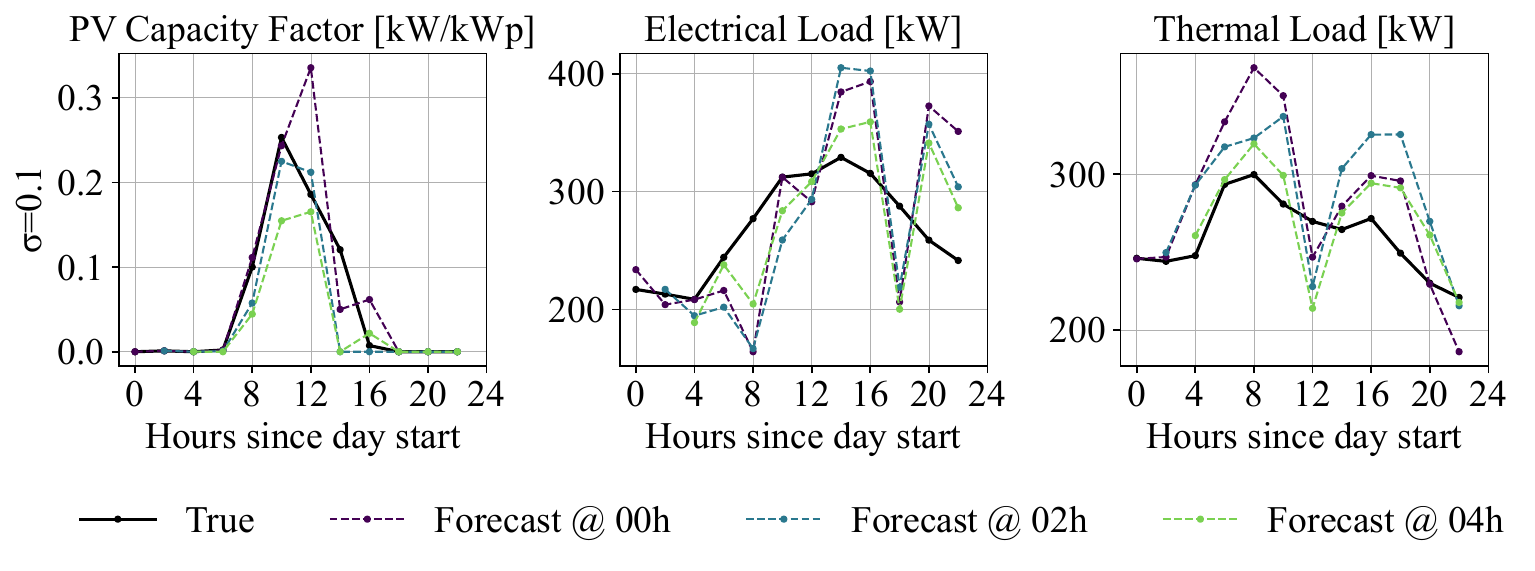}
    \caption{Consecutive forecasts constructed from the MMFE samples plotted over a typical day for the PV capacity factor, electrical demand, and heat demand, with a fixed $\sigma=0.10$ and $\rho=0.78$.}
    \label{fig:disturbances}
\end{figure}

\begin{figure}[t]
    \centering
    \includegraphics[width=1.0\linewidth]{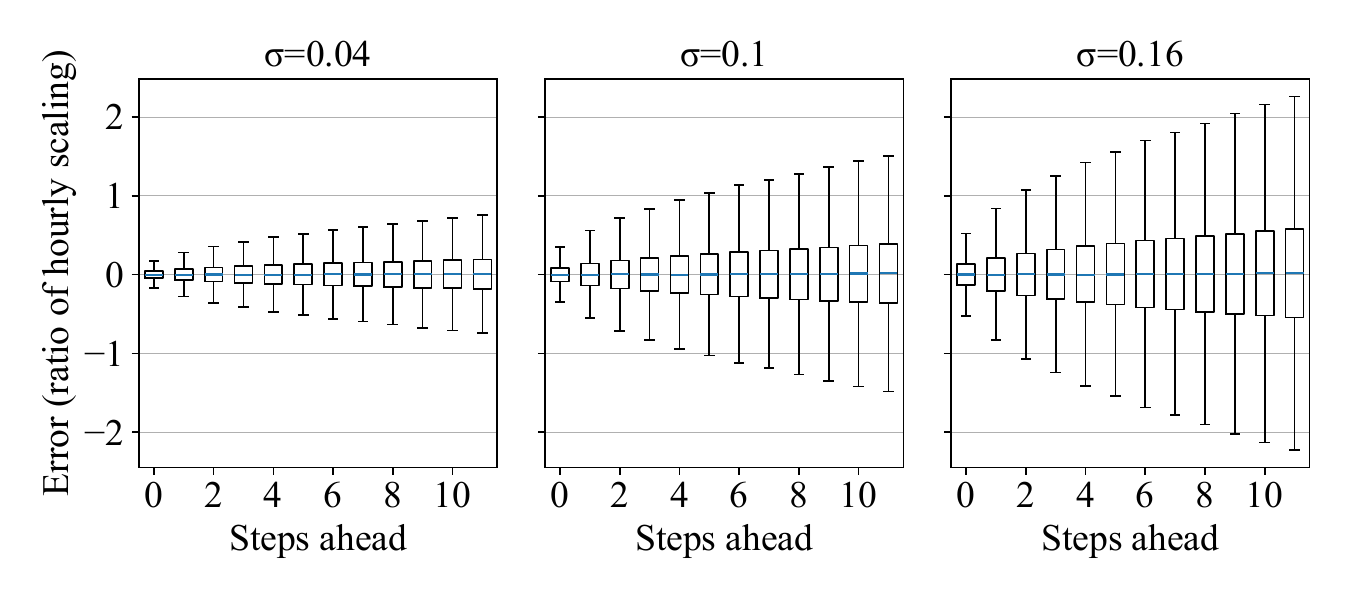}
    \caption{Forecast error, normalized by the hourly scaling value in \eqref{eq:mmfe_error_scaled}, versus lead time for different levels of $\sigma$ with $\rho$ = 0.78.}
    \label{fig:forecast_errors}
\end{figure}

We compare single-level sizing of energy system devices by solving \eqref{opt:open_loop_design} with bi-level sizing via BP-MPC (Algorithm \ref{alg:BP-MPC}) across forecast error parameters $\sigma$ and $\rho$, evaluating device capacities and annualized costs under deterministic MPC. We denote the single-level design with open-loop operation by SL-OLC and its subsequent MPC operation by SL-MPC. Both methods use the same realized disturbances for sizing and evaluation, while BP-MPC additionally uses the MMFE forecasts supplied to the MPC. Costs are reported relative to the SL-OLC objective, i.e., the cost anticipated by single-level planning, and differences between designs in percentage points (pp) of it. Reported objectives exclude the regularization term \eqref{eqn:reg}, which contributes less than 0.1\% of the SL-OLC objective in all cases. 

We use 2024 demand and weather data from ETH Z\"urich's ETZ building at a $2$-hour resolution, with an annual planning horizon ($T=4380$) and a $24$-hour MPC horizon ($N=12$). Economic parameters are listed in Table \ref{tab:hub_prices} and unless stated otherwise, $c^{\mathrm{viol}}=3$ CHF/kWh and $\rho=0.78$, which corresponds to approximately 8 hours of correlated forecast updates. The grid applies the high tariff from 6 a.m. to 10 p.m. on Monday through Saturday and the low tariff otherwise. Energy system parameters and algorithm settings are provided in the Appendix. Optimization problems are implemented in Python using CasADi \cite{andersson_casadi_2019}, solving the MPC QPs with qpOASES \cite{ferreau_qpoases_2014} and solving the single-level QP with Gurobi \cite{gurobi_optimization_llc_gurobi_2026}.

\begin{table}[t]
  \centering
  \caption{Energy system prices} 
  \label{tab:hub_prices}
  \small
  \begin{tabular}{@{}llc@{}}
    \toprule
    Type & Parameter & Value \\
    \midrule
    
    Investment & $c^{\mathrm{inv}}_{\mathrm{pv}}$, $c^{\mathrm{inv}}_{\mathrm{hp}}$, $c^{\mathrm{inv}}_{\mathrm{bess}}$, $c^{\mathrm{inv}}_{\mathrm{tes}}$ & 61, 205, 40, 3.5 CHF/unit/yr\\
    
    Grid & $\underline{c}^{\mathrm{buy}}$, $\overline{c}^{\mathrm{buy}}$, $c^{\mathrm{sell}}$& 0.22, 0.27, 0.0 CHF/kWh\\    
    \bottomrule
  \end{tabular}
\end{table}

Figure \ref{fig:disturbances} illustrates the MMFE forecasts for $\sigma=0.1$ and $\rho=0.78$, showing that errors shrink as the target timestep approaches and successive forecasts remain coherent while still allowing random revisions. For the PV capacity factor, the lead-time effect is dominated by the diurnal scaling. Figure \ref{fig:forecast_errors} shows the normalized, unclipped errors across the horizon for several values of $\sigma$, illustrating how larger $\sigma$ widens the error distributions at all lead times.

\begin{figure}
    \centering
    \includegraphics[width=1.0\linewidth]{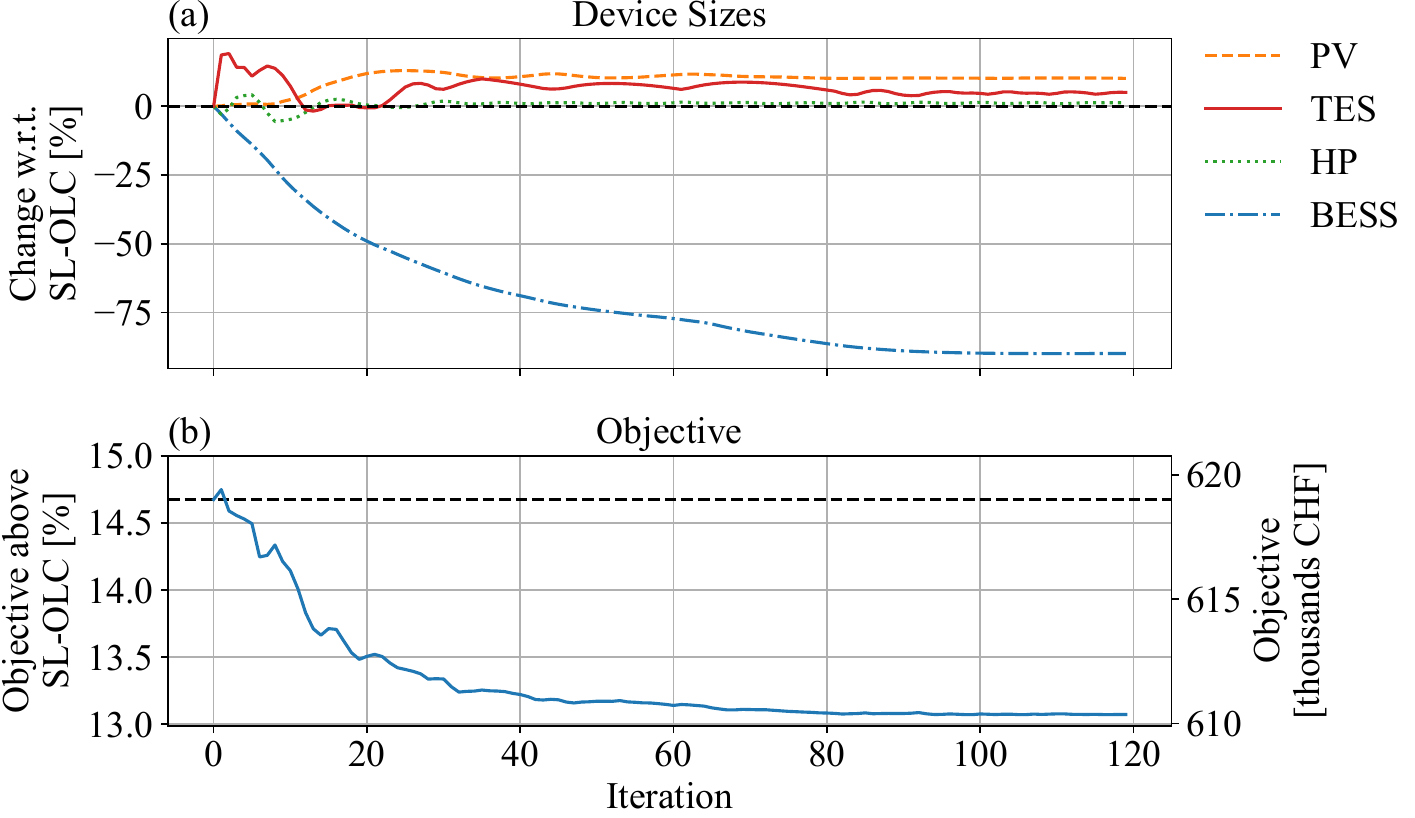}
    \caption{Convergence of BP-MPC from the SL-OLC design for $\sigma = 0.10$ and $\rho = 0.78$. The dashed line marks (a) the SL-OLC design and (b) its cost under MPC (SL-MPC).}
    \label{fig:convergence}
\end{figure}

\begin{figure}[t]
    \centering
    \includegraphics[width=1.0\linewidth]{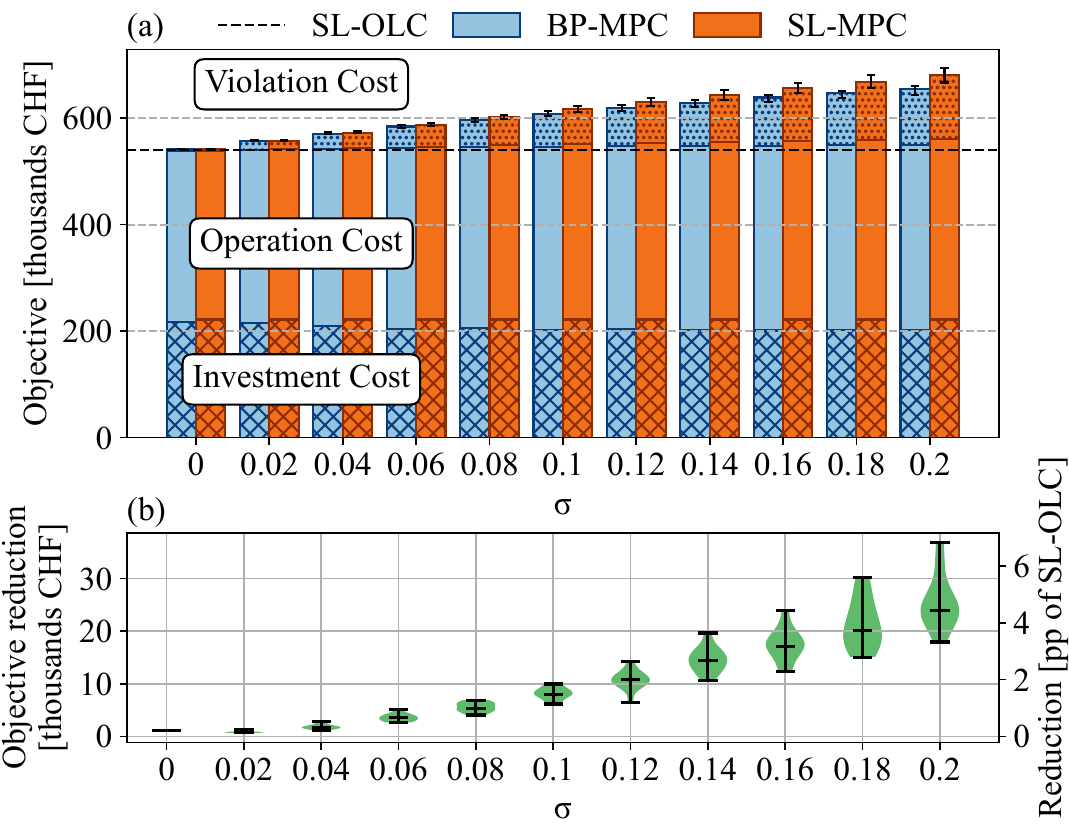}
    \caption{Effect of $\sigma$ on the energy system cost over the 11 forecast scenarios. (a) Mean cost breakdowns for designs operated with MPC, with error bars showing the range. (b) Cost advantage of BP-MPC, with violin markers showing the minimum, median and maximum.}
    \label{fig:objective_sigma_sweep}
\end{figure}

\begin{figure}[b]
    \centering
    \includegraphics[width=1.0\linewidth]{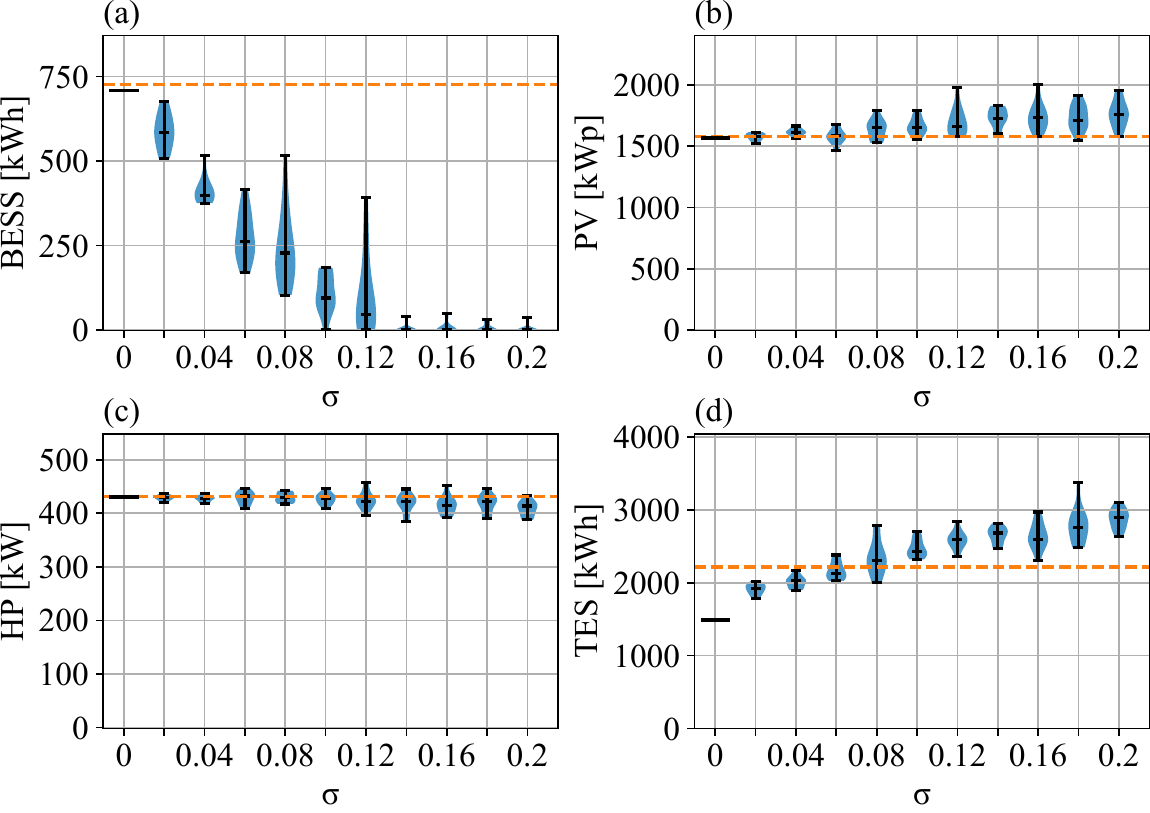}
    \caption{Effect of $\sigma$ on the BP-MPC device sizes over the 11 forecast scenarios (blue violin plots). The dashed orange line is the SL-OLC design.}
    \label{fig:devices_sigma_sweep}
\end{figure}

\begin{figure}[t]
    \centering
    \includegraphics[width=1.0\linewidth]{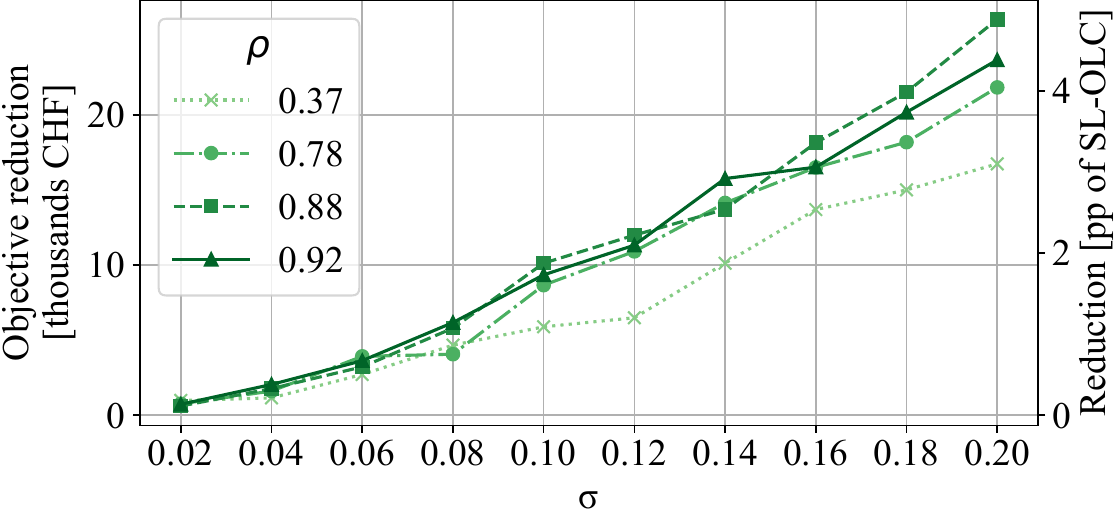}
    \caption{Objective reduction of BP-MPC across $\sigma$ for different $\rho$, corresponding to correlated-error time spans of 2 to 24 hrs, for a single annual forecast error scenario. Darker colors indicate higher $\rho$.}
    \label{fig:sig_rho_sweep}
\end{figure}

\begin{figure}[b]
    \centering
    \includegraphics[width=0.9\linewidth]{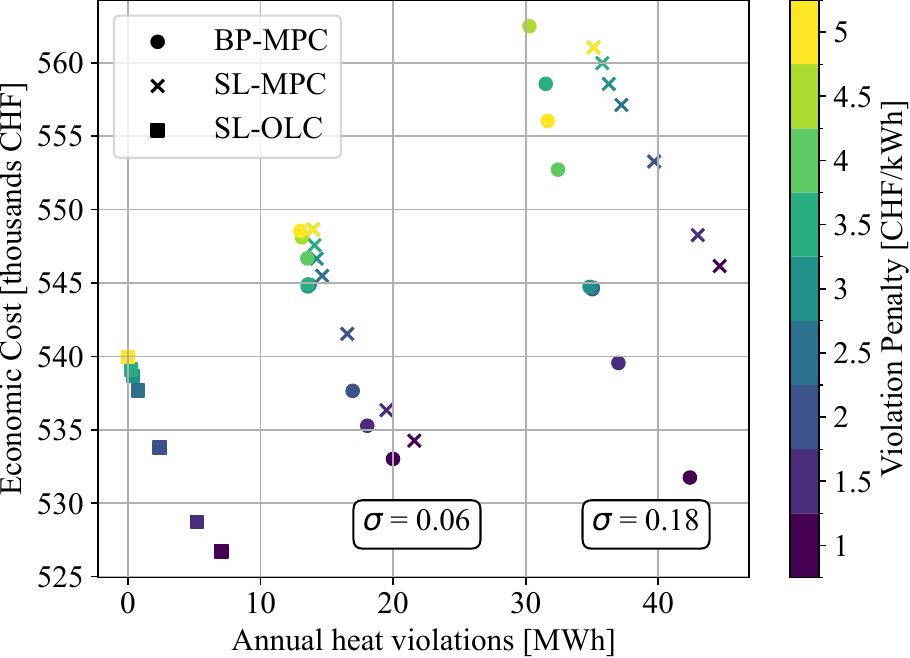}
    \caption{Trade-off between annual economic cost and heat violations from varying violation penalties at different values of $\sigma$. For reference, the annual heat demand is just over $1500$ MWh.}
    \label{fig:viol_penalty_sweep}
\end{figure}

Figure \ref{fig:convergence} depicts the convergence of BP-MPC for one annual forecast scenario with $\sigma=0.1$ and $\rho=0.78$, highlighting three differences. First, the initial single-level design costs 14.7\% more under deterministic MPC than estimated by SL-OLC as evident in Figure \ref{fig:convergence}b. Second, BP-MPC settles near a fixed point after about 100 iterations, yielding a much smaller BESS and slightly larger PV and TES capacities as shown in Figure \ref{fig:convergence}a. The lower BESS investment suggests that forecast errors reduce the value of battery-based load shifting. For example, overestimating surplus PV generation can cause insufficient charging during low-tariff hours or an over-reliance on midday PV charging, which increases high-tariff grid purchases. Third, these design changes lower the closed-loop cost to 13.1\% above the SL-OLC estimate, a reduction of 1.6 pp.

Extending the comparison of BP-MPC to the single-level benchmark across values of $\sigma$ shows that these differences become more pronounced as forecast error increases. Figures \ref{fig:objective_sigma_sweep} and \ref{fig:devices_sigma_sweep} show system costs and designs for 11 annual forecast scenarios at each $\sigma$, with BP-MPC optimizing a separate design tailored to each scenario. The resulting savings are therefore in-sample and represent an upper bound for a single design applied to unseen forecasts. Figure \ref{fig:objective_sigma_sweep}a illustrates that the SL-OLC estimate increasingly understates the closed-loop costs of both designs as forecast error grows, mainly due to the cost of rising heat violations. BP-MPC nevertheless improves on the single-level design in every evaluated case, with average annual cost savings increasing from near zero at $\sigma=0$ to over 4 pp of the SL-OLC cost at $\sigma=0.2$ as presented in Figure \ref{fig:objective_sigma_sweep}b. The cost breakdown in Figure \ref{fig:objective_sigma_sweep}a shows that BP-MPC achieves these savings by trading lower investment for a slight increase in operating costs, while keeping violation costs similar or lower. Overall, these results indicate that both the costs overlooked by single-level estimates and the benefits of optimizing for closed-loop operation increase with forecast error.

Figure \ref{fig:devices_sigma_sweep} shows that the BP-MPC capacities increasingly diverge from the single-level design to attain their savings as forecast error grows. As $\sigma$ increases, BP-MPC progressively reduces BESS capacity, eliminating it in most scenarios for $\sigma \geq 0.14$, while increasing PV and TES capacities. Note that without forecast error, BP-MPC selects a smaller TES than the single-level design because of the controller's limited horizon. TES capacity then increases with the forecast error to exceed the single-level capacity. The overall investment shift supports the earlier interpretation that forecast errors reduce BESS value, favoring investment in PV and TES instead under the evaluated MPC policy. Additional PV capacity provides more energy, while a larger TES supports heat load shifting and buffers forecast mismatches. These insights into cost savings and device investment for energy systems design emerge from the bi-level formulation, which treats the limitations of operational forecasts explicitly, in contrast to single-level formulations.

Figures \ref{fig:sig_rho_sweep} and \ref{fig:viol_penalty_sweep} show that the growing advantage of BP-MPC with increasing $\sigma$ also holds across error correlations and heat-violation penalties. Both sweeps use one annual forecast-error sample, evaluated under the indicated parameter values. Figure \ref{fig:sig_rho_sweep} shows that the BP-MPC benefit is generally smaller for weakly correlated errors ($\rho = 0.37$) than for $\rho \geq 0.78$, particularly at higher $\sigma$.
Higher $\rho$ makes sequences of over- or underprediction more persistent, which can disrupt storage scheduling and lead to more heat violations and grid purchases. The modest impact is consistent with $\rho$ changing the persistence of errors while leaving their one-step marginal distribution unchanged. Ultimately, the trend of greater BP-MPC savings at larger $\sigma$ holds for all evaluated values of $\rho$.

Figure \ref{fig:viol_penalty_sweep} shows the trade-off between heat violations and economic cost obtained by varying the heat-violation penalty at two forecast-error levels. BP-MPC attains a lower objective than SL-MPC for every penalty, mostly by reducing heat violations, although it can trade slightly more violations for a lower economic cost at low penalties. The performance gap between BP-MPC and SL-MPC is larger at $\sigma=0.18$ than at $\sigma=0.06$ for every penalty, and this difference exceeds its variation across penalties. This further supports the previously observed trend of greater BP-MPC benefits with increasing $\sigma$, which remains dominant across the evaluated penalties.

The remaining heat violations in Figure \ref{fig:viol_penalty_sweep} for both values of $\sigma$ despite high violation penalties demonstrate the importance of the control policy. Although the TES can buffer forecast mismatches, deterministic MPC leaves insufficient flexibility to accommodate unexpected shifts in heat demand when it schedules the storage to be empty or full. Therefore, the ability of capacity adjustments to reduce heat violations is limited. If preventing most or all heat violations is a design requirement for the energy system, this can motivate implementing controllers that explicitly account for forecast uncertainty, such as stochastic or robust MPC. Ultimately, the achievable benefit of bi-level planning depends on the rollout policy, reinforcing the importance of considering closed-loop behavior during planning.

\section{Conclusion}
We proposed a bi-level optimization method for sizing PV, BESS, HP, and TES capacities for an energy system based on BP-MPC, a gradient-based method that accounts for closed-loop MPC operation under limited, evolving forecasts. We compared the resulting designs with conventional single-level sizing, which assumes unlimited foresight, and quantified how the benefit of bi-level sizing depends on forecast accuracy. Across the forecast error levels evaluated using a parameterized MMFE model, bi-level sizing reduced costs relative to single-level sizing in all evaluated scenarios, achieving larger average benefits at higher forecast error levels. A single-scenario sweep showed the same pattern across different forecast error correlations and violation penalties. Increasing forecast error also shifted investment away from BESS toward PV and TES. The absolute gains are modest in our case study because the deterministic MPC does not hedge against forecast errors. Larger gains are expected where the controller's limited foresight matters more, for example with seasonal storage or peak-based tariffs, whose value depends on information far beyond the forecast horizon. Future work will extend the analysis to such settings using suitable terminal ingredients and uncertainty-aware controllers, such as stochastic MPC, whose parameters can be co-designed with the device sizes.




\setcounter{table}{0}
\renewcommand{\thetable}{A\arabic{table}}
\section*{APPENDIX}
This appendix provides additional simulation details. The energy system model parameters are listed in Table \ref{tab:hub_params}, and we set $\underline{p}=0.1$ units for each device. In the regularization cost \eqref{eqn:reg}, the scaling matrices $S_x$ and $S_u$ normalize the states and inputs by the maximum allowed device sizes $\overline{p}$. We heuristically set $\gamma=0.2$ and $d_j=\Delta t$ for terms $j$ that penalize BESS and TES charging and discharging. BP-MPC uses $\beta=0.9$, gradient scaling by $D=\mathrm{diag}(c^{\mathrm{inv}})$, norm clipping at $r=3$, and step sizes $\alpha_i=0.075(1+0.1i)^{-0.51}$. Instead of a stopping criterion, we run a fixed budget of 120 iterations, which allows convergence to be verified visually (see Figure \ref{fig:convergence}).
\begin{table}[h!]
  \centering
  \caption{Energy system model parameters}
  \label{tab:hub_params}
  \small
  \begin{tabular}{@{}llc@{}}
    \toprule
    Type & Parameter & Value \\
    \midrule
    HP & $\eta_\mathrm{COP}$
        & 4.5\\

    BESS  & $\eta^\mathrm{bess}$, $\eta^\mathrm{bess}_\mathrm{ch}$, $\eta^\mathrm{bess}_\mathrm{dc}$,&0.999, 0.95, 0.95,\\ 
    &$\underline{a}^\mathrm{bess}$, $\overline{a}^\mathrm{bess}$, $\overline{b}^{\mathrm{bess}}$ 
        & 0.2, 0.9, 0.25\\
        
    TES  & $\eta^\mathrm{tes}$,
    $\eta^\mathrm{tes}_\mathrm{ch}$,
    $\eta^\mathrm{tes}_\mathrm{dc}$,& 0.992, 0.99, 0.99,\\ 
    &$\underline{a}^\mathrm{tes}$,
    $\overline{a}^\mathrm{tes}$,
    $\overline{b}^{\mathrm{tes}}$ 
    & $10^{-6}$, 1.0, 0.25\\

    Capacity limits & $\overline{p}^\mathrm{pv}$, $\overline{p}^\mathrm{hp}$, & 2000 kWp, 1000 kW,\\  
    &$\overline{p}^\mathrm{bess}$, $\overline{p}^\mathrm{tes}$,&1000 kWh, 5000 kWh\\
    \bottomrule
  \end{tabular}
\end{table}

\section*{ACKNOWLEDGMENT}

This work was supported as a part of NCCR Automation, a National Centre of Competence in Research, funded by the Swiss National Science Foundation (grant number 51NF40\_225155). Claude Opus 5.5 and GPT 6.1 Sol were used for code feature expansion and for pre-submission feedback, including suggested text edits and references. The research content, results, and conclusions are the authors’ own; all code, text, and references were reviewed, revised, and approved by the authors.

\balance
\bibliography{references}

\end{document}